# Beyond Cognitive Load:

## AI-Based Estimation of Cognitive Effort Using Brain Signals During Digital Tasks

Shayla Sharmin, MSc; Mohammad Fahim Abrar, MD; Gael Lucero-Palacios; Aditya Raikwar; Roghayeh Leila Barmaki

Computer and Information Sciences, University of Delaware

## Abstract

**Background.** Cognitive effort, defined as the relationship between cognitive load and task performance, offers insight into how individuals efficiently allocate mental resources during cognitively demanding activities. This metric is crucial in high-stakes public health and clinical training, where unmanaged cognitive overload has been linked to medical errors and workforce burnout. This study aims to examine whether cognitive effort varies systematically across task segments and whether it can be estimated at the individual level using brain signal data and machine learning. **Method.** Functional near-infrared spectroscopy (fNIRS) data were collected from 16 participants during a structured digital cognitive task comprising four sequential segments separated by short and long rest intervals. Cognitive effort was defined through relative neural efficiency and relative neural involvement, which combined measures of prefrontal hemodynamic activity with task performance. The analysis followed a two-stage approach. First, a segment-level group analysis assessed whether cognitive effort differed significantly across predefined task segments, thereby confirming that the task structure produced meaningful changes in cognitive demand. Second, participant-independent machine learning models predicted task performance from brain signal features. These predicted performance scores were then combined with neural measures to estimate cognitive effort at the individual level. **Results.** First, statistical analysis showed significant differences in cognitive effort across the four task segments. This confirms that even small changes in the assessment structure impact on the collective cognitive efficiency of the trainees. Then, we used machine learning on fNIRS data to predict individual performance scores. We found that the effort calculated from these predicted scores was nearly identical to that calculated from the actual scores. This result suggests that our effort metric is strongly based on brain signals. **Conclusion.** The findings demonstrate the feasibility of estimating cognitive effort from brain signals using artificial intelligence at both group and individual levels. **Public Health Implications.** Estimation of cognitive effort from digital task data may support scalable monitoring of cognitive workload and mental fatigue in technology-mediated environments, which complements subjective assessment methods used in public and digital health research.

## Introduction

Digital training and assessment environments are now common in public health and clinical settings, but they often rely on observable performance metrics to infer competence and readiness.[1] Performance alone does not reveal the cognitive resources required to achieve that performance, and unmanaged cognitive overload has been associated with adverse outcomes such as errors and burnout.[2,3] Objective approaches that quantify cognitive effort during technology-mediated tasks can strengthen training design and support a safer, more resilient public health workforce. Accuracy and response time are widely used indicators in interactive digital tasks, yet they provide an incomplete and sometimes inconsistent view of internal cognitive state. Two individuals can reach the same score with very different mental effort, and low performance can reflect overload, lapses in attention, or strategy differences rather than disengagement,[4,5] which are critical risk factors in clinical practice.

Cognitive Load Theory explains performance constraints through limited working memory capacity and distinguishes between productive and unproductive load.[6–8]However, load estimates based on behavior or self-report alone cannot reliably capture real-time effort investment during task execution.[9–11] Task performance and neural activity, when considered independently, provide an incomplete and sometimes ambiguous characterization of an individual's cognitive state during task execution. To address this limitation, cognitive effort can be operationalized by combining neural measures of cognitive load with task performance. Getchell and Shewokis proposed Relative Neural Efficiency (RNE) and Relative Neural Involvement (RNI) to represent cognitive effort to distinguish efficient engagement, overload, disengagement, and deep involvement.[5,12,13] RNE reflects the efficiency with which an individual achieves task performance relative to cognitive load, whereas RNI reflects the degree of engagement and involvement during task execution.[3,14] Together, these two metrics provide a comprehensive view of efficiency and involvement, which offers actionable indicators for optimizing learning strategies and improving cognitive performance.

Prior work has examined cognitive effort at the group level[3,15,16] or used machine learning to predict task outcomes,[17–19] but fewer studies connect validated segment dynamics with individual-level effort estimation within a single framework. In this study, we adopt a two-level analytical approach. We first conducted a segment-level analysis in which cognitive effort was examined across predefined task segments by aggregating data across participants. At the individual level, cognitive effort is estimated for each participant by integrating machine learning–predicted performance with brain signal measures, which enables personalized effort estimation beyond observed task outcomes. We address the following research questions:

**(RQ1) Segment-Level Validation:** How does cognitive effort (measured by RNE and RNI) evolve across consecutive task segments, and are the changes statistically significant at the group level?

**(RQ2) Individual-Level Estimation:** Can individual cognitive effort be accurately estimated by integrating machine learning-predicted performance scores into the effort calculation model?

To address these questions, we conducted a controlled study using functional near-infrared spectroscopy to measure prefrontal brain activity during a structured digital cognitive task and applied machine learning to estimate cognitive effort at individual levels. This research makes the following key contributions:

**Two-Stage Framework for Cognitive Effort Estimation:** We introduce a two-stage framework that first validates segment-level variations in cognitive effort across the group and then estimates individual-level cognitive effort using predicted performance derived from brain signals.

**Analysis of Effort Dynamics:** We demonstrate how cognitive effort evolves across task segments with different rest structures, which highlights systematic changes in efficiency and involvement over time.

**Implications for Cognitive Workload Monitoring:** This work provides a scalable foundation for objective monitoring of cognitive workload in technology-mediated tasks, with potential relevance for digital training, assessment, and public health research contexts.

The goal of this work is not to optimize machine learning prediction accuracy. Instead, machine learning is used to estimate individual cognitive effort when ground-truth performance is unavailable. The central contribution of this study is to first statistically determine the significance of cognitive effort (RNE/RNI) across systematic task segments and subsequently validate the feasibility of personalized RNE/RNI estimation for individuals using neurophysiological data and predicted task outcomes. This approach emphasizes the validation of the effort framework at the single-subject level, rather than the development of a high-performing prediction model.

## Related Work

### Cognitive Effort: Relative Neural Efficiency and Involvement

In public health and clinical environments, understanding the mental energy investment, or **Cognitive effort**, which a learner uses to complete a task, is crucial for preventing critical errors. Cognitive effort refers to the relationship between task performance and the cognitive load an individual invests in performing a task.[3,5,12] Cognitive load reflects the demands placed on working memory by task complexity and instructional design.[7,9–11] However, cognitive load alone does not indicate how much effort a learner actually expends. For example, a novice may exert high effort on a simple task, whereas an expert may complete a complex task with relatively low effort. Functional near-infrared spectroscopy (fNIRS) provides an objective way to examine cognitive effort by measuring changes in prefrontal hemodynamic activity associated with mental demand.[20] To combine neural activity with task outcomes, prior work introduced two complementary metrics: Relative Neural Efficiency (RNE) and Relative Neural Involvement (RNI).[5] As illustrated in Figure 1, RNE reflects how efficient performance is achieved relative to cognitive load, while RNI captures the degree of engagement or effort investment during task execution.

Figure 1 Conceptual Plots Illustrating Cognitive Effort.

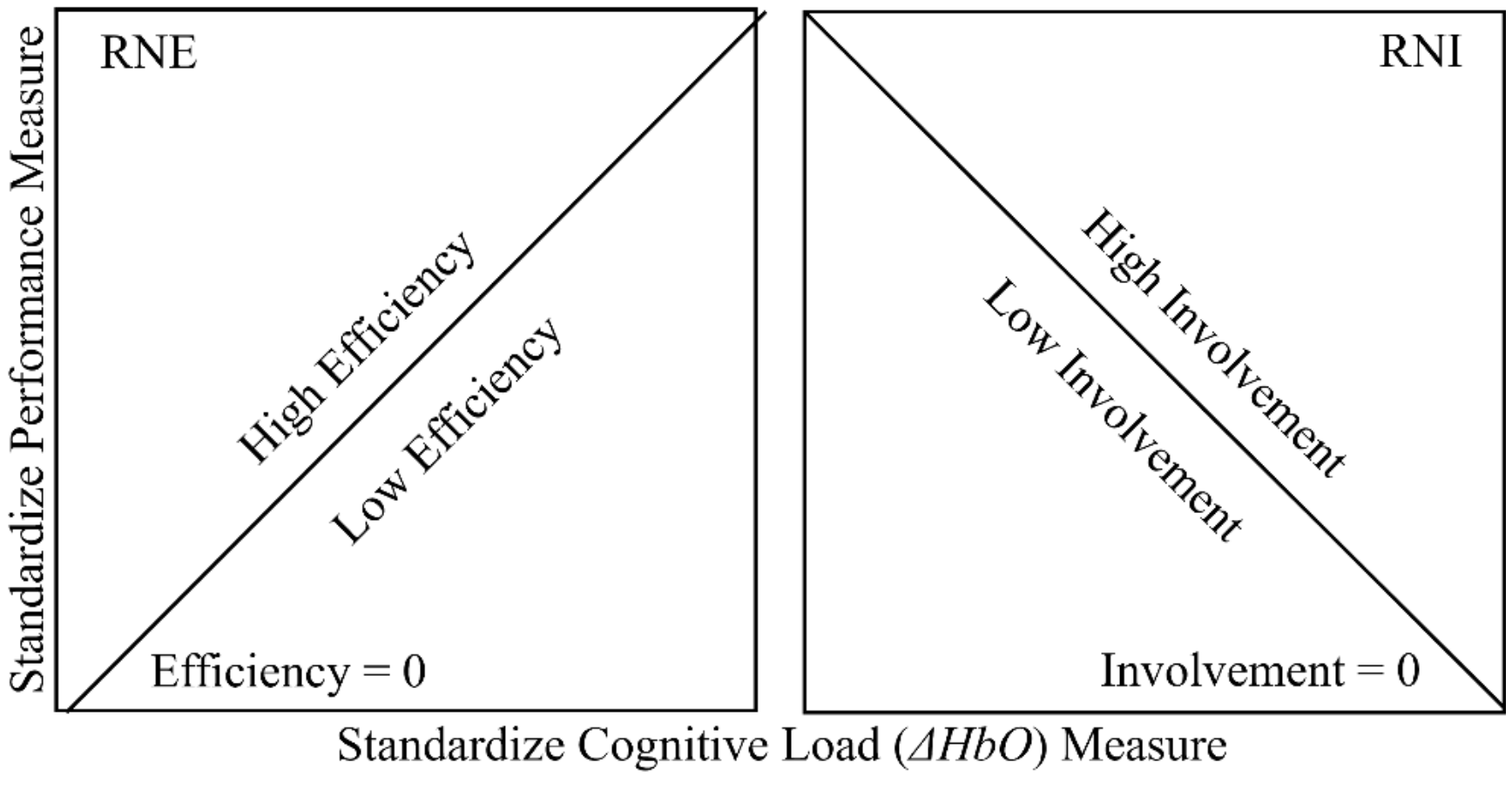

The left plot shows Relative Neural Efficiency (RNE), where higher performance relative to cognitive load indicates high efficiency. The right plot shows Relative Neural Involvement (RNI), where higher cognitive load relative to performance indicates high involvement. The X-axis represents standardized cognitive load, and the Y-axis represents standardized performance.

Studies comparing virtual and physical task environments have demonstrated systematic differences in neural efficiency attributable to task modality. Other work has shown that prior experience, such as extensive gaming or skill training, is associated with higher neural efficiency during cognitively demanding tasks.[16] Similar group-level effects have been reported in domains including practice-based learning, brain stimulation paradigms, and surgical simulation training.[14] Cognitive effort metrics are sensitive even to subtle variations in task execution. For example, differences in interaction modality during a quiz-based learning task (e.g., hand input versus stylus input) produced measurable changes in cognitive effort.[3] These findings highlight that RNE and RNI capture fine-grained changes in mental resource allocation that may not be apparent from performance metrics alone.

However, these studies primarily rely on post-hoc, group-level statistical analysis. While they reveal meaningful trends (e.g., experts are more efficient than novices), they do not provide a framework for estimating cognitive effort at the individual level. To the best of our knowledge, prior work has not used machine learning predicted performance to compute RNE and RNI for individual learners. This limitation motivates the need for methods that enable individualized, scalable cognitive effort estimation without relying on ground-truth performance scores.

## Machine Learning for Cognitive State and Performance Prediction

Machine learning techniques have increasingly been applied to fNIRS data to support automated, scalable interpretation of brain activity in real-world settings. These approaches are particularly valuable in contexts where continuous human monitoring is impractical, such as digital training, assessment, and operational environments relevant to public health and clinical practice.

Across a broad range of applications, machine learning models have been used to decode neural signals associated with task execution, interaction, and cognitive control. Prior work has demonstrated successful classification of speech-related activity,[21] virtual environment interaction,[22] motor imagery,[23] and fine motor control[24] from fNIRS signals. Multimodal approaches combining EEG and fNIRS have further improved robustness and generalizability in brain–computer interface systems.[25]

Beyond interaction tasks, machine learning has been widely applied to identify cognitive states that are directly relevant to safety, performance, and well-being. In complex operational environments, models such as convolutional and recurrent neural networks have been used to predict cognitive and perceptual load during prolonged or high-demand tasks, including flight simulations and control scenarios.[26,27] Other studies have shown that simpler models, such as logistic regression, can reliably classify workload levels in standardized task-load simulations.[28] Mental fatigue prediction has also received growing attention, with evidence linking sustained prefrontal hemodynamic changes to fatigue accumulation during extended task performance.[29,30]

In learning and training contexts, machine learning has been used to estimate task performance, engagement, and cognitive difficulty from fNIRS data. These studies demonstrate that neural signals collected during task execution can predict behavioral outcomes and reflect changes in neural efficiency over time.[18,19] However, most existing approaches focus on predicting performance or workload as isolated outcomes. While clustering and classification methods have been explored, generalization across individuals remains challenging due to inter-subject variability in brain responses.[31]

Taken together, prior work establishes that machine learning can extract meaningful cognitive information from fNIRS signals across diverse task domains. However, existing studies typically treat cognitive load, performance, or fatigue as separate targets. Few approaches integrate predicted performance with neural measures to estimate cognitive effort as a unified construct, and none, to our knowledge, apply this integration to derive individual-level measures such as relative neural efficiency and involvement.

## Methods

This section describes the study design, data collection procedures, and analytical methods used in this research. We first outline the experimental protocol and the participant's interaction with a structured digital cognitive task during which brain activity was recorded. Next, we describe how cognitive effort was operationalized and quantified at both the segment and individual levels using fNIRS signals and task performance data. Finally, we detail the analytical approach used to estimate individual cognitive effort by predicting task outcomes using machine learning models. This highlights how artificial intelligence supports objective and scalable assessments relevant to public health and digital training contexts.

### Study Design

In this study, participants completed a structured digital cognitive task designed to elicit varying levels of cognitive effort during problem solving. The task was implemented in a gamified format to maintain engagement while systematically manipulating cognitive demand across task segments, as illustrated in Figure 2.

Throughout the task, participants' performance outcomes were recorded, and prefrontal hemodynamic responses were continuously measured using fNIRS. Demographic information and pre- and post-task responses were collected using the Qualtrics survey platform.

Figure 2 Overview of the Study Procedures

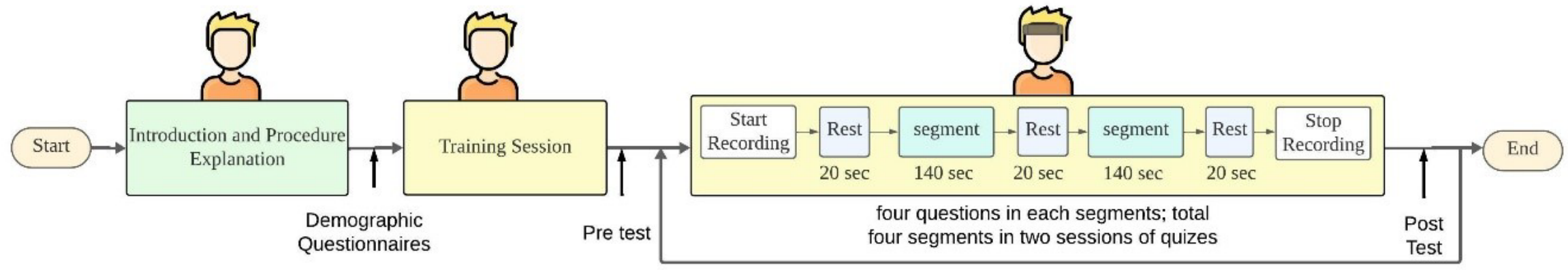

Participants began with consent, demographic questionnaires, and a demonstration of the task. Each session started with a 20-second resting baseline, followed by 140-second task segments containing four questions each.

The task used in this study was based on a simple graph representation consisting of nodes and edges. Graph-based structures are used in public health and biomedical research to model complex relationships, making them an ecologically relevant yet domain-neutral task structure.[32] This provides a controlled relational framework that elicits abstraction, reasoning, and working memory demands, making them suitable for examining cognitive effort under standardized digital assessment conditions. The study protocol was reviewed and approved by the institutional review board, and all participants provided informed consent before participation.

## Study Procedure

After providing informed consent and completing a demographic questionnaire, participants were familiarized with the study protocol through a brief demonstration session. Instructions emphasized minimizing head and body movement to reduce motion artifacts during brain signal acquisition. Participants then completed a ten-item pre-test assessing baseline understanding of the task material. This was followed by a standardized instructional video introducing the task concepts and response format.

The main task consisted of 16 questions organized into two sessions, with each session comprising two segments of four questions. Participants were given up to 30 seconds to respond to each question, followed by 5 seconds of performance feedback. Each segment, therefore, lasted 140 seconds, consisting of 120 seconds of task execution and 20 seconds of feedback. A 20-second rest period separated the two segments within each session.

A longer rest interval of approximately 6–10 minutes was provided between the two sessions to mitigate fatigue. After each session, participants completed a brief post-test. Question order was randomized across participants to minimize learning effects and control for task difficulty bias.

## Participants

An a priori power analysis was conducted using G*Power.[33] A medium effect size was assumed (Cohen’s f = 0.25, corresponding to partial eta squared $\eta^2_p = 0.06$), with $\alpha = 0.05$, statistical power of 0.80, and a sphericity correction factor of $\varepsilon = 0.75$.

The analysis indicated a minimum sample size of 16 participants. This target sample size accounted for potential data loss due to participant dropout and fNIRS signal quality issues. Healthy graduate students between the ages of 23 and 32 years were recruited for the study. All participants provided informed consent before participation. Exclusion criteria included sensitivity to alcohol-based cleaning solutions used during sensor preparation and prior professional experience with graph-based tasks within the past five years, to minimize the effects of content familiarity on performance and cognitive effort estimation.

A total of 20 participants were enrolled in the study (Table 1). Four participants were excluded from analysis: two due to fNIRS sensor malfunction and two due to atypical hemodynamic signal patterns identified as statistical outliers using the Interquartile Range method applied to hemodynamic response value. The final dataset consisted of 16 participants (11 female, 5 male, mean age = 27.13 ± 2.5 years).

Table 1. Demographic Summary of Study Participants (N = 16).

| Characteristics | Count | % |
|---|---|---|
| **Gender** | | |
| Female | 11 | 68.75% |
| Male | 5 | 31.25% |
| **Education** | | |
| Graduate program | 16 | 100% |
| **Prior digital task experience** | | |

| Characteristics | Count | % |
|---|---|---|
| Never | 9 | 56.25% |
| A few times per week | 3 | 18.75% |
| A few times per year | 4 | 25.00% |
| **Task content familiarity** | | |
| Somewhat confident | 1 | 6.25% |
| Neutral | 3 | 18.75% |
| Somewhat not confident | 4 | 25.00% |
| Not confident | 8 | 50.00% |

The table reports counts and percentages for gender, education level, prior exposure to digital tasks, and self-reported familiarity with task content.

Baseline demographic characteristics, prior exposure to game-based tasks, and self-reported familiarity with the task content were collected to characterize factors that could influence task performance and cognitive effort estimates. Prior exposure to game-based digital tasks was collected to characterize baseline familiarity with interactive digital task formats that could influence task performance and cognitive effort estimates.

## Apparatus

Figure 3 illustrates the experimental setup and fNIRS sensor configuration (a). Participants completed the structured digital cognitive task on a laptop computer, while a separate desktop computer was used for fNIRS data acquisition and monitoring. Brain activity was recorded using a continuous-wave functional near-infrared spectroscopy system (Imager 2000S, fNIR Devices LLC, Potomac, MD, USA) equipped with an 18-channel headband sensor pad. The sensor array consisted of four light-emitting diode sources operating at wavelengths of 730~nm and 850~nm, and ten photodetectors, with a source–detector separation of 2.5~cm.

Figure 3. Experimental Setup

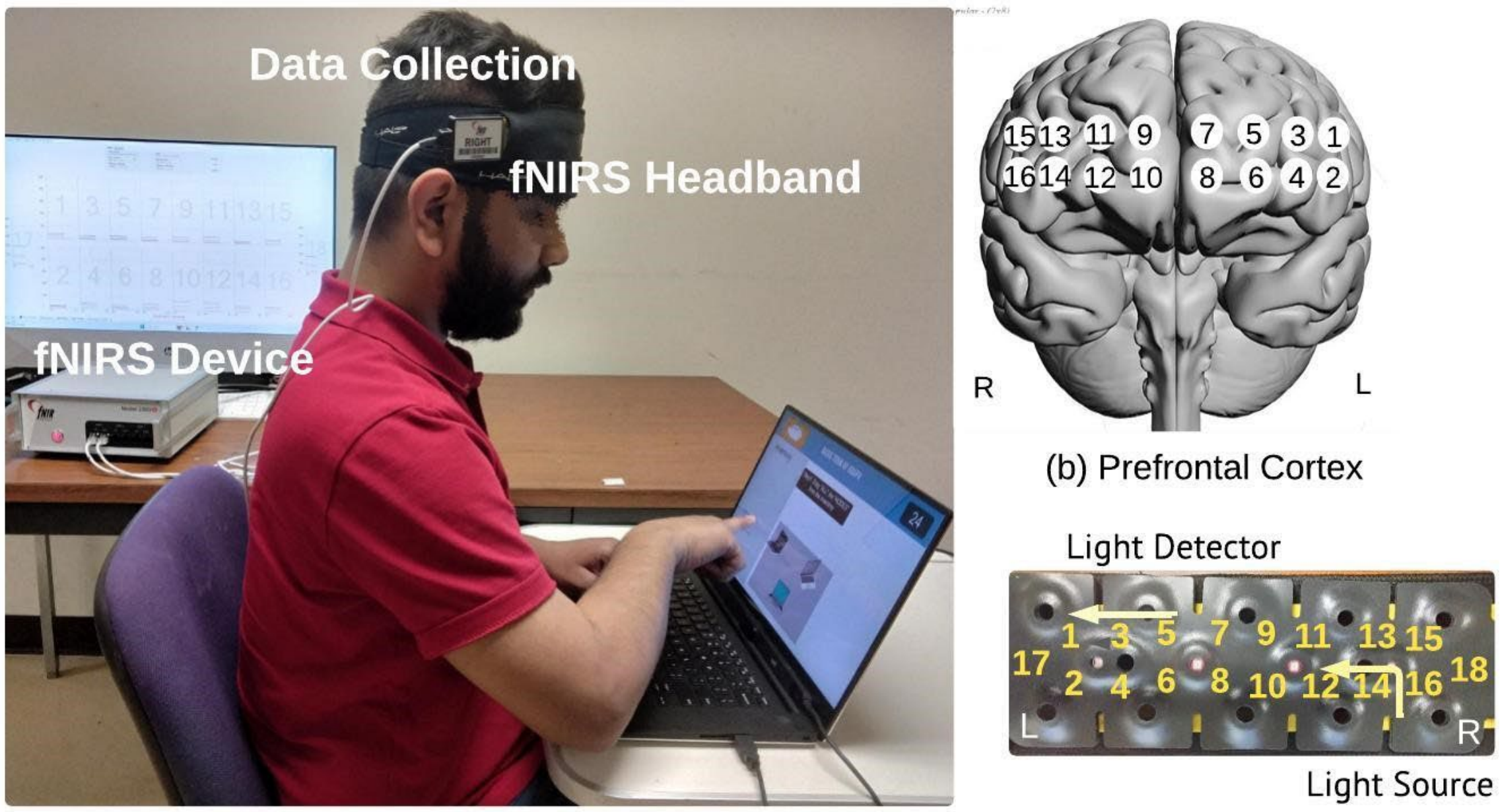

(a) Study Setup

(c) fNIRS Headband

The sensor array was positioned over the prefrontal cortex (PFC), with channels 1–16 covering regions corresponding to Brodmann Areas 9, 10, 44, and 45 (b). Sensor placement followed the International 10–20 system, with the center of the headband aligned to the Fpz location and the horizontal axis extending toward Fp1 and Fp2.[11,34,35] Two additional channels (channels 17 and 18), located on the lateral extensions of the sensor pad (c), served as reference channels to capture systemic physiological signals, such as scalp blood flow, rather than cortical hemodynamic activity. We divided the segment by sending biomarker from a python code.[36,37]

Data acquisition was performed using the Cognitive Optical Brain Imaging Studio software,[38] and preprocessing was conducted using fNIRSoft (version~4.9).[39]

## Signal Processing

Raw fNIRS signals were preprocessed using fNIRSoft software (version~4.9). Preprocessing steps were applied to improve signal quality and reduce physiological and instrumental noise before analysis. First, signal quality was assessed by visually inspecting the raw light-intensity data from all optodes. Channels exhibiting poor contact or excessive attenuation, often due to hair obstruction or sensor misalignment, were excluded from further analysis.

Next, physiological noise was attenuated using a finite impulse response (FIR) low-pass filter (20th order, Hamming window) with a cutoff frequency of 0.1~Hz to reduce high-frequency components associated with cardiac and respiratory activity. To address slow baseline drift, a linear detrending procedure was applied to remove low-frequency trends from the signal without introducing phase distortion. Filtered light intensity signals were then converted into changes in oxygenated (ΔHbO) and deoxygenated hemoglobin (ΔHbR) concentrations using the Modified Beer–Lambert Law.[34]

To reduce the influence of systemic physiological signals unrelated to cortical activity, a reference-channel-subtraction approach was employed. Signals from the two lateral reference channels were averaged to estimate global systemic fluctuations and subtracted from the 16 prefrontal channels. This procedure minimized contamination from extracerebral sources, such as scalp blood flow, allowing subsequent analyses to focus on task-related hemodynamic responses.

## Operationalization of Cognitive Effort

Cognitive effort was operationalized using **Relative Neural Efficiency (RNE)** and **Relative Neural Involvement (RNI)**, which together characterize the relationship between task performance and cognitive load. Cognitive load was indexed using mean changes in oxygenated hemoglobin (ΔHbO), while task performance was quantified using quiz scores. RNE reflects the efficiency with which task performance is achieved relative to neural activation, whereas RNI reflects the degree of engagement and voluntary effort exerted during task execution.

## Standardization of Performance and Cognitive Load

To enable meaningful comparison between neural activity and task performance, both performance scores and ΔHbO values were standardized using *z*-scores. For each measure, the mean (μ) and standard deviation (σ) were computed, and standardized values were calculated using the following formula:

$$z = \frac{(x - \mu)}{\sigma}$$

In cases where the standard deviation was zero (i.e., no variability within a segment), all standardized values were set to zero to avoid undefined values.

Standardization served two purposes: (1) reducing inter-individual variability, and (2) placing physiological and behavioral measures on a common scale suitable for combined analysis. In the equations below, $P_z$ denotes the standardized performance score, and $M_z$ denotes the standardized cognitive load derived from ΔHbO.

## Segment-Level Cognitive Effort (Aggregated Across Participants)

For segment-level analysis, cognitive effort was examined across predefined task segments using data aggregated across participants. For each segment, performance scores $P_{Si}$ and cognitive load values $M_{Si}$ from all participants were pooled to compute segment-specific means and standard deviations. Standardized values were calculated as follows:

$$P_z = \frac{\left(Ps_{\mathrm{i}} - \mu(Ps)\right)}{\sigma(Ps)} \quad (1)$$

$$M_z = \frac{\left(Ms_{\mathrm{i}} - \mu(Ms)\right)}{\sigma(Ms)} \quad (2)$$

To align with established interpretations of neural efficiency, ΔHbO values were inversely transformed such that lower prefrontal activation corresponded to greater neural efficiency. This approach is consistent with prior literature, which interprets reduced neural activation during correct task performance as indicative of more efficient neural processing.

### Individual-Level Cognitive Effort (Within-Participant Estimation)

For individual-level estimation, cognitive effort was computed separately for each participant using their own data across the four task segments. For participant i and segment s, mean performance and mean cognitive load were first calculated. Standardization was then performed within each participant using participant-specific means and standard deviations:

$$Pz(s,i) = \frac{\bar{P}(s,i) - \mu(\bar{P}i)}{\sigma(\bar{P}i)} \quad (3)$$

$$Mz(s,i) = \frac{\bar{M}(s,i) - \mu(\bar{M}i)}{\sigma(\bar{M}i)} \quad (4)$$

This approach expresses cognitive effort for each segment relative to an individual's own baseline, preserving inter-individual differences while enabling personalized estimation of cognitive effort.

### Computation of Relative Neural Efficiency and Involvement

Using the standardized performance ($P_z$) and cognitive load ($M_z$) values, Relative Neural Efficiency and Relative Neural Involvement were computed using a Cartesian transformation:

$$RNE = \frac{(P_z - M_z)}{\sqrt{2}} \quad (5)$$

$$RNI = \frac{(P_z + M_z)}{\sqrt{2}} \quad (6)$$

For segment-level analysis, $P_z$ and $M_z$ corresponded to standardized values aggregated across participants for each task segment. For individual-level analysis, $P_z$ and $M_z$ were replaced by participant-specific standardized values. This unified computational framework enabled direct comparison of cognitive effort patterns at both segment and individual levels.

### Estimating Cognitive Effort using Machine Learning

The following sections describe how brain signal data were prepared and analyzed using machine learning to estimate individual task performance, which serves as an intermediate step for personalized cognitive effort assessment in technology-mediated public health and training contexts (see Figure 4).

Figure 4: Overview of the Machine Learning Pipeline

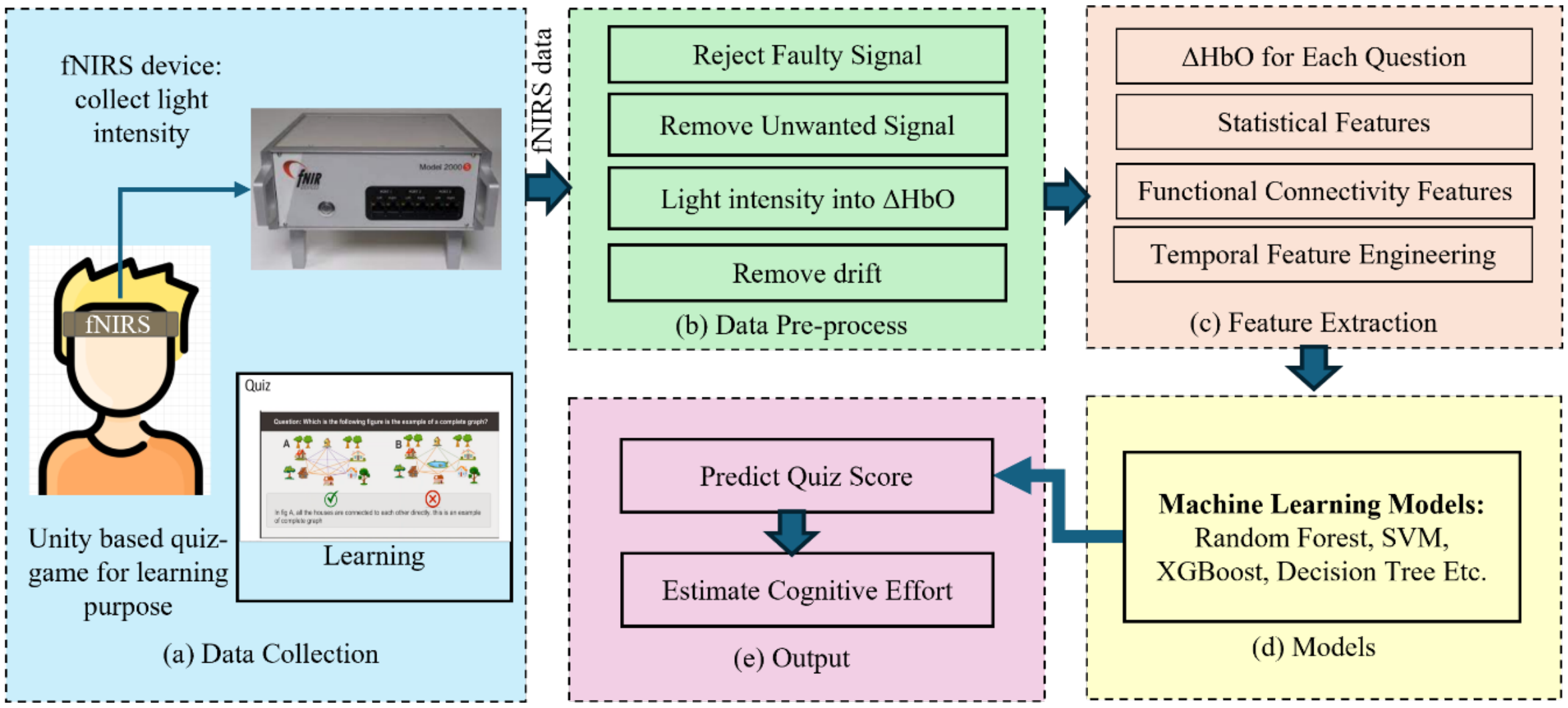

(a) fNIRS data are collected while participants play an educational quiz game. (b) Signals are pre-processed, including noise removal and conversion of light intensity to hemodynamic response. (c) Hemodynamic response is extracted for each question, and statistical, functional connectivity, and temporal features are derived. (d) Machine learning models (e.g., Random Forest, SVM, XGBoost, Decision Tree) are used to predict quiz scores. (e) Predicted scores and hemodynamic response are combined to estimate cognitive effort through relative neural efficiency and involvement.

## Data Preparation

After signal preprocessing, fNIRS data were segmented at the question level. Each question trial originally consisted of 300 time points, corresponding to 30 seconds of recording at a sampling rate of 10 Hz. Behavioral inspections showed that most participants responded within 20 seconds. To maintain temporal consistency across trials and avoid including post-response noise, only the first 200 data points (20 seconds) were retained for each question. Each of the 16 participants completed 16 questions, resulting in 256 question-level samples. Each sample contained time-series data from 16 prefrontal optodes, yielding a total of 819,200 data points across the dataset.

Task performance was treated as a binary outcome, with correct responses labeled as Class 1 and incorrect responses labeled as Class 0. The dataset contained 168 correct and 88 incorrect responses, reflecting a natural class imbalance commonly observed in cognitive task performance data. Given the limited sample size and the goal of preserving ecological validity, no synthetic oversampling techniques were applied. Instead, model evaluation emphasized Precision, Recall, and F1-score rather than overall accuracy, ensuring sensitivity to incorrect responses, which are more informative for cognitive effort estimation.

## Machine Learning for Performance Estimation

Machine learning was used as an intermediate step to estimate individual task performance from brain signal data. The primary objective was not to maximize prediction accuracy, but to determine whether predicted performance derived from fNIRS signals could be reliably used to compute individual-level cognitive effort. This step is critical for enabling objective effort estimation when behavioral outcomes are incomplete, delayed, or unavailable.

**Feature Construction:** Feature vectors were constructed using oxygenated and deoxygenated hemoglobin signals extracted from prefrontal channels. All features were standardized using z-score normalization to ensure comparable scaling across participants and recording sessions.

**Classification Models:** To assess robustness across modeling approaches, multiple supervised classification algorithms were evaluated, including Logistic Regression, Support Vector Machines with a radial basis function kernel, K-Nearest Neighbors (k=5), Linear Discriminant Analysis, Decision Trees, Random Forests, eXtreme Gradient Boosting, and Naive Bayes. Each model produced a binary prediction indicating whether task performance was correct or incorrect.

**Participant-Independent Validation:** To ensure generalizability beyond individual participants, a 5-fold Group K-Fold cross-validation strategy was employed. This approach ensured that data from any given participant appeared exclusively in either the training or testing set within a fold, preventing data leakage. In each fold, approximately 80% (12 or 13) of participants were used for training, and the remaining 20% (3 or 4) were reserved for testing.

**Evaluation Metrics:** Model performance was evaluated using Precision, Recall, and F1-score rather than overall accuracy. This choice reflects the study's emphasis on reliable detection of incorrect responses, which carry greater informational value for cognitive effort estimation than correct responses alone.

## Results

Before estimating cognitive effort at the individual level, we first examined whether the task structure produced systematic and statistically significant variations in cognitive effort across predefined task segments. This analysis served as a validation step to confirm that segment-wise changes in cognitive load and performance reflected meaningful differences in cognitive state rather than random variability.

### Segment-Level Validation of Cognitive Effort

Figure 5 illustrates the group-level cognitive effort trajectory across the four task segments in Cartesian space, using standardized cognitive load $M_z$ and performance $P_z$. Each point represents the average cognitive state across participants within a segment. Relative Neural Efficiency (RNE) and Relative Neural Involvement (RNI) were derived as projections onto the efficiency (Y=X) and involvement (Y=-X) axes, respectively.

Figure 5. Cartesian plot of Standardized Cognitive Load (X-axis) and Standardized Performance (Y-axis) Across Four Quiz Sessions.

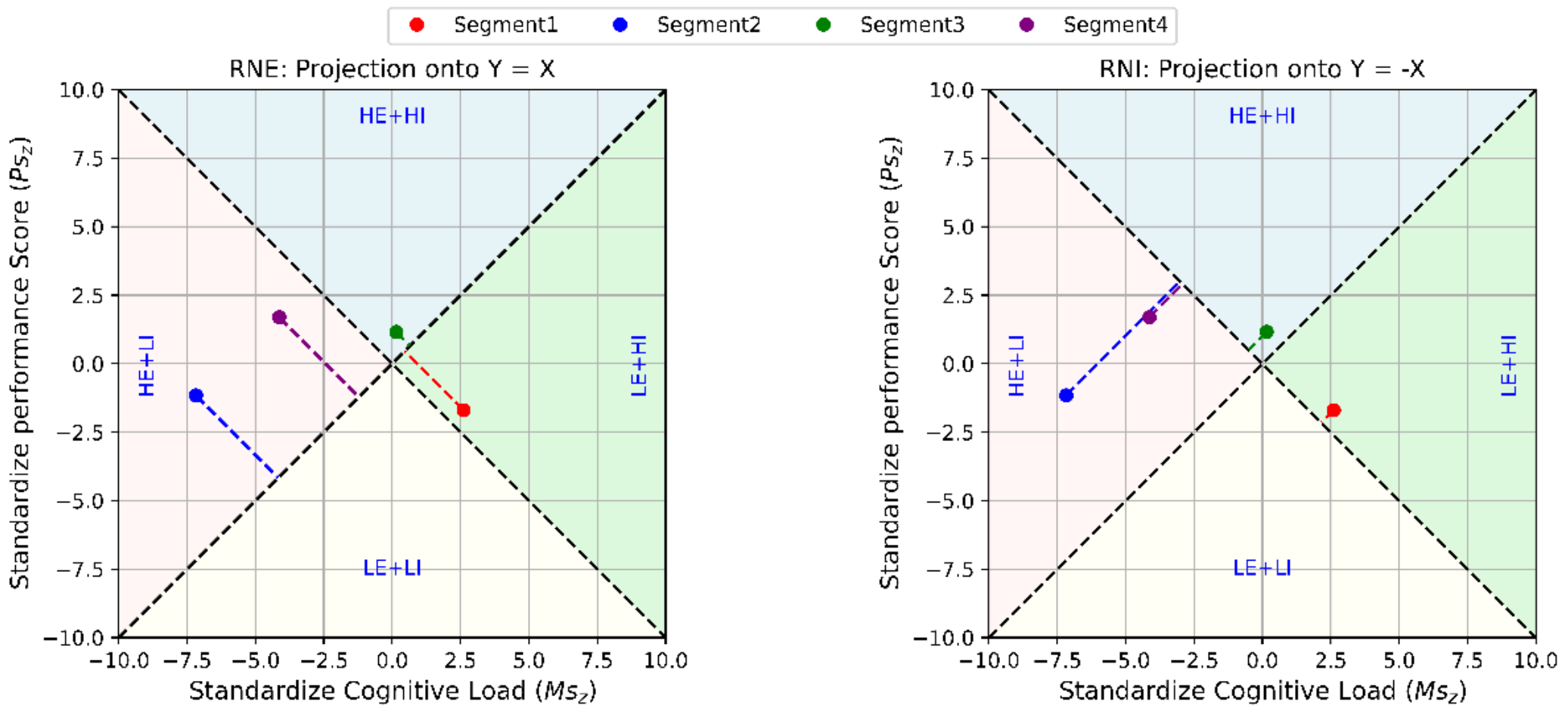

Each session's dot reflects the average cognitive state during that segment. The dashed diagonal lines represent the efficiency axis (Y = X) and involvement axis (Y = -X). Relative Neural Efficiency (RNE) is computed as the projection onto the Y=X line, and Relative Neural Involvement (RNI) onto Y=-X. Arrows illustrate segment-to-segment transitions. Segment 1 shows high involvement but low efficiency; Segment 2 exhibits low involvement despite high efficiency; Segment 3 reflects re-engagement after a longer break; and Segment 4 shows improved performance with minimal effort.

Distinct segment-wise patterns were observed. Segment 1 was characterized by high cognitive load and relatively low performance, resulting in low RNE and high RNI, indicative of high effort with limited efficiency. Following a short rest period, Segment 2 showed a marked reduction in cognitive load without a corresponding improvement in performance, yielding higher RNE but lower RNI. After a longer break, Segment 3 demonstrated moderate cognitive load and slightly improved performance, with near-neutral efficiency and involvement. Segment 4 maintained performance with reduced cognitive load, resulting in higher efficiency and lower involvement.

Statistical comparisons confirmed that these segment-wise changes were significant. Wilcoxon signed-rank tests revealed significant increases in RNE from Segment 1 to Segment 2 ($p = 0.04$, $r = 0.72$), a decrease from Segment 2 to Segment 3 ($p < 0.001$, $r = -0.45$), and an increase from Segment 3 to Segment 4 ($p < 0.005$, $r = 0.25$) (see Figure 6).

Figure 6 Mean Relative Neural Efficiency (RNE) across four task segments for all participants.

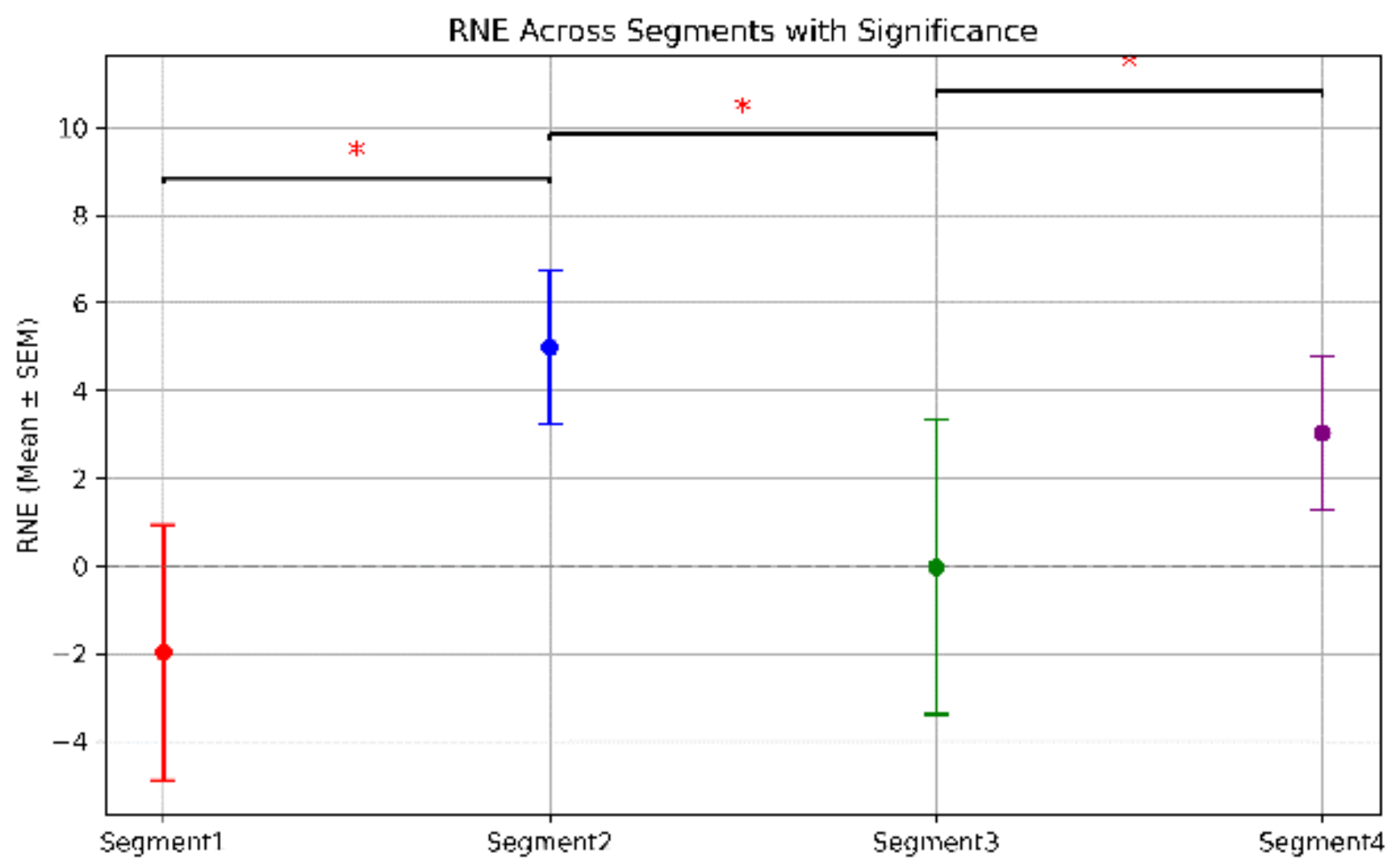

Each point represents the mean value for a segment, and vertical error bars indicate the standard error of the mean (SEM). Red asterisks denote statistically significant pairwise differences identified using Wilcoxon tests ($p < 0.05$).

RNI showed a significant decrease from Segment 1 to Segment 2 ($p = 0.02$, $r = -0.74$), an increase from Segment 2 to Segment 3 ($p = 0.01$, $r = 0.48$), and a smaller decrease from Segment 3 to Segment 4 ($p = 0.03$, $r = -0.24$) (see Figure 7).

Figure 7 Mean Relative Neural Involvement (RNI) across four task segments.

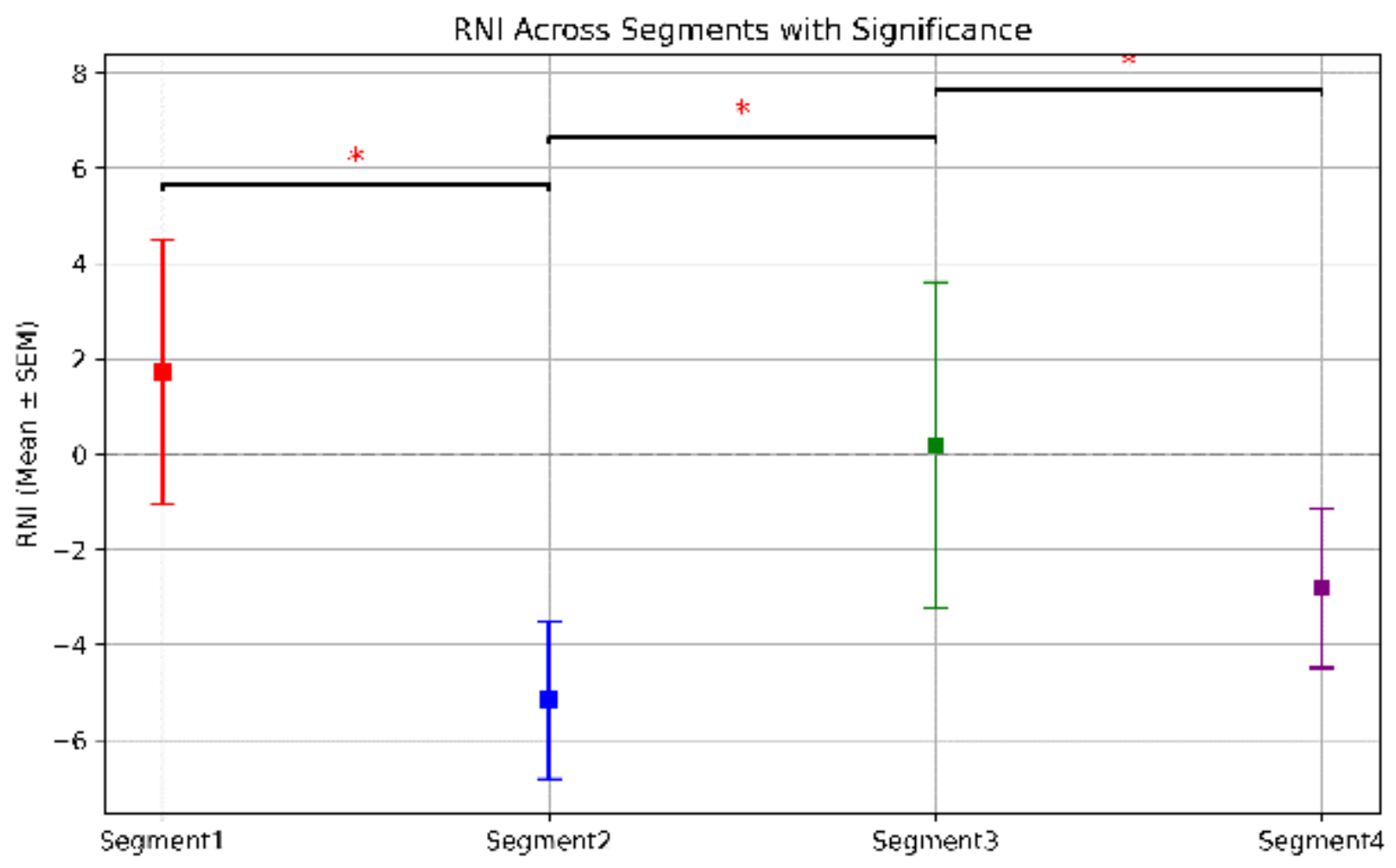

Each point represents the mean value for a segment, and vertical error bars indicate the standard error of the mean (SEM). Red asterisks denote statistically significant pairwise differences identified using Wilcoxon tests ($p < 0.05$).

Together, these results demonstrate that cognitive effort, as quantified by RNE and RNI, varies systematically across task segments. This confirms that the task design induces meaningful and measurable changes in cognitive state, supporting its suitability for subsequent individual-level cognitive effort estimation.

### Individual-Level Estimation of Cognitive Effort Using Predicted Performance (RQ2)

Table 2 presents the performance of multiple machine learning models across three feature configurations: oxygenated hemoglobin (ΔHbO), deoxygenated hemoglobin (ΔHbR), and their combination. The dataset showed a natural class imbalance, with approximately 65.6% correct responses. A naive majority-class baseline that always predicts "correct" would therefore achieve an accuracy of 0.66. However, such a baseline fails to identify incorrect responses and yields zero Precision, Recall, and F1-score for the minority class. As a result, it is not suitable for estimating cognitive effort.

Table 2. Performance Evaluation of Machine Learning Models Across Different Feature Sets

| Feature Set | Model | Accuracy | Precision | Recall | F1 Score |
|---|---|---|---|---|---|
| **ΔHbO Only** | Decision Tree | 0.57 | 0.57 | 0.57 | 0.56 |
| | KNN (k = 5) | 0.62 | 0.60 | 0.62 | 0.57 |
| | LDA | 0.65 | 0.52 | 0.65 | 0.55 |
| | Logistic Regression | 0.65 | 0.53 | 0.65 | 0.56 |
| | Naive Bayes | 0.65 | 0.53 | 0.65 | 0.57 |
| | Random Forest | 0.60 | 0.57 | 0.60 | 0.57 |
| | SVM (RBF) | 0.65 | 0.47 | 0.65 | 0.54 |
| | XGBoost | 0.59 | 0.57 | 0.59 | 0.56 |
| **ΔHbR Only** | Decision Tree | 0.59 | 0.62 | 0.59 | 0.59 |
| | KNN (k = 5) | 0.58 | 0.57 | 0.58 | 0.55 |
| | LDA | 0.63 | 0.52 | 0.63 | 0.55 |
| | Logistic Regression | 0.65 | 0.57 | 0.65 | 0.57 |
| | Naive Bayes | 0.65 | 0.61 | 0.65 | 0.58 |
| | Random Forest | 0.60 | 0.56 | 0.60 | 0.56 |
| | SVM (RBF) | 0.64 | 0.47 | 0.64 | 0.53 |
| | XGBoost | 0.61 | 0.62 | 0.61 | 0.60 |
| **Combined** | Decision Tree | 0.53 | 0.55 | 0.53 | 0.54 |
| | KNN (k = 5) | 0.61 | 0.63 | 0.61 | 0.57 |
| | LDA | 0.66 | 0.62 | 0.66 | 0.61 |
| | **Logistic Regression** | **0.67** | **0.63** | **0.67** | **0.61** |
| | Naive Bayes | 0.65 | 0.53 | 0.65 | 0.57 |
| | Random Forest | 0.61 | 0.56 | 0.61 | 0.57 |
| | SVM (RBF) | 0.64 | 0.47 | 0.64 | 0.54 |
| | XGBoost | 0.62 | 0.63 | 0.62 | 0.59 |

Values are reported as mean scores from a 5-fold Group K-Fold cross-validation. The highest F1 score for each feature set is highlighted in bold.

Across evaluated models, linear classifiers using combined ΔHbO and ΔHbR features showed the most balanced performance. Logistic Regression and Linear Discriminant Analysis achieved the highest F1-scores (approximately 0.61), indicating that task performance can be reasonably estimated from fNIRS signals. Although the improvement in overall accuracy over the naive baseline was modest, these models successfully distinguished between correct and incorrect responses rather than defaulting to majority-class predictions.

The purpose of this analysis was not to achieve high score prediction accuracy, but to determine whether predicted performance scores are sufficient for estimating individual cognitive effort.

### Individual-Level Estimation of Cognitive Effort

For each participant, the quiz was divided into four segments, each consisting of four consecutive questions. Within each segment, cognitive load was computed using standardized fNIRS-derived measures, and performance was represented using either the actual score or the predicted score. Relative Neural Efficiency (RNE) and Relative Neural Involvement (RNI) were then calculated for each segment using the same formulation for both actual and predicted performance.

Figure 8 compares RNE and RNI values computed using actual scores versus predicted scores for four representative participants. Across participants, RNE showed stronger alignment between actual and predicted values than RNI. For example, Participant 16 exhibited near one-to-one correspondence between actual and predicted RNE values. Other participants, such as Participants 10 and 8, showed greater variability, particularly for RNI.

Figure 8. Scatter Plots Comparing Actual Versus Predicted Cognitive Effort Metrics for Four Participants (Test Case)

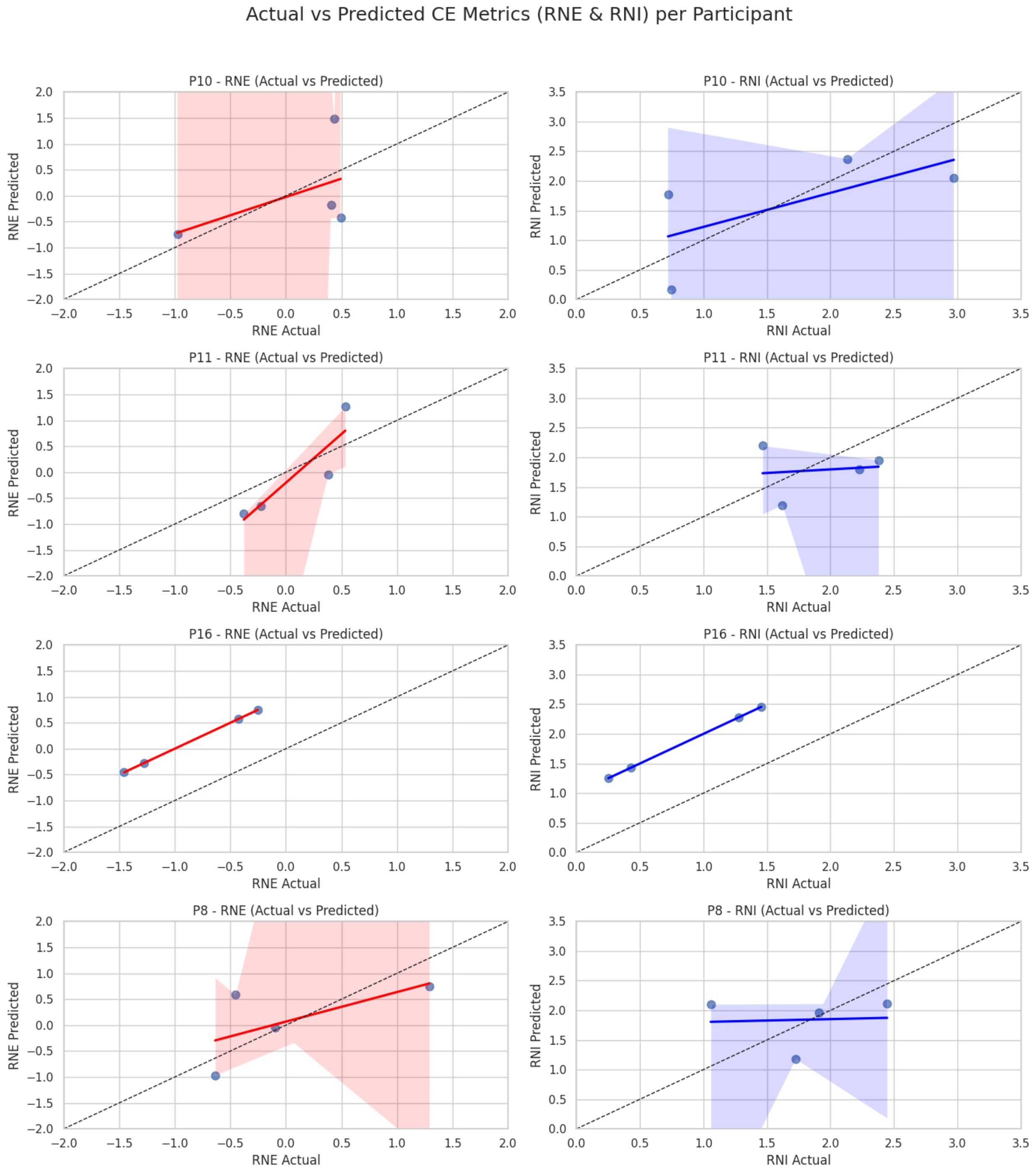

The left column shows Relative Neural Efficiency (RNE), and the right column shows Relative Neural Involvement (RNI). Each point represents one data segment for a participant. The dashed diagonal lines indicate perfect agreement between actual and predicted values. Red lines (RNE) and blue lines (RNI) represent the regression fit for the model, showing how closely predictions align with actual data.

Despite variability at the individual prediction level, the overall patterns of cognitive effort were preserved. These findings indicate that machine learning–predicted performance captures the general structure of cognitive effort, even when exact score prediction is imperfect.

### Statistical Validation of Estimated Cognitive Effort

To evaluate whether predicted and actual cognitive effort values differed systematically, Wilcoxon signed-rank tests were applied. No significant differences were observed between actual and predicted values for either RNE ($p = 0.211$) or RNI ($p = 0.211$). This result indicates that predicted and actual cognitive effort metrics follow comparable distributions. Correlation analysis showed that RNE had a slightly stronger association ($r = 0.51$) with performance than RNI ($r = 0.45$). Both metrics showed identical mean absolute error (MAE = 0.67) and mean squared error (MSE = 0.56). Although these differences were modest, they suggest that efficiency-based measures may be more stable than involvement-based measures under the conditions of this study.

## Discussion

This study presents a two-stage framework for estimating cognitive effort during digital learning tasks. In the first stage, we examined whether relative neural efficiency and relative neural involvement, derived from fNIRS signals and actual task performance, captured meaningful and statistically significant patterns at the group level across consecutive task segments.

In the second stage, we evaluated whether cognitive effort could be estimated at the individual level by integrating machine learning–predicted performance with brain signal data. This approach allowed us to assess whether predicted scores could substitute for actual performance without distorting cognitive effort metrics.

By linking brain signals with predicted performance, the proposed framework provides a scalable foundation for interpreting cognitive effort in technology-mediated learning environments. Importantly, these findings address a gap in prior work, where cognitive load and performance were often examined separately, and relative neural efficiency and involvement were primarily limited to group-level analyses.

### Cognitive Effort Patterns Across Consecutive Task Segments (Group-Level Analysis) (RQ1)

The segment-level analysis addressed the first research question by establishing that cognitive effort varies systematically across consecutive task segments. This finding shows that the observed effort metrics reflect structured changes in cognitive state. The observed pattern aligns with theoretical models of cognitive adaptation, in which individuals initially invest high effort to manage unfamiliar task demands and gradually transition toward more efficient processing. Early segments were characterized by high involvement and low efficiency, which suggests substantial mental resource investment with limited performance gains. Subsequent segments showed shifts toward greater efficiency and reduced involvement, consistent with task familiarization and cognitive adjustment.

From a public health perspective, these results have important implications. Training and assessment tasks in healthcare and public health settings are often structured into repeated segments separated by brief or extended rest periods. The present findings demonstrate that such structural features can significantly influence cognitive effort, even when performance outcomes appear stable. Monitoring cognitive effort at the segment level may therefore provide early indicators of cognitive overload, disengagement, or fatigue that are not captured by performance metrics alone.

By validating that segment-wise cognitive effort changes are both systematic and statistically significant, this analysis provides a necessary foundation for individual-level cognitive effort estimation. Establishing this baseline ensures that subsequent machine learning–based predictions of individual effort are grounded in task-induced cognitive dynamics rather than incidental variability.

## Individual Cognitive Effort Estimation Using Predicted Performance (RQ2)

The second research question showed whether individual cognitive effort can be estimated using machine-learning–predicted performance rather than actual scores. While classification accuracy was moderate, the results demonstrate that predicted performance scores are sufficient to preserve individual cognitive effort patterns.

A key finding is that substituting predicted scores for actual scores did not significantly alter the resulting RNE and RNI values. Although both actual and predicted effort calculations share the same cognitive load component, the critical test was whether replacing actual performance would distort the cognitive effort structure. The absence of significant differences between actual and predicted effort distributions confirms that this substitution is statistically valid.

Relative Neural Efficiency proved to be more robust than Relative Neural Involvement across participants. This suggests that efficiency-based measures, which reflect how effectively cognitive resources are translated into performance, may be more reliable indicators of cognitive state than involvement-based measures, which may capture additional motivational or affective factors that are harder to estimate from brain signals alone.

From a public health perspective, these findings are important because they support the feasibility of estimating cognitive effort without relying on explicit performance feedback. In digital training and assessment environments, such as those used in public health and clinical workforce preparation, objective monitoring of cognitive workload and fatigue may be possible even when behavioral outcomes are delayed, incomplete, or unavailable. This capability provides a foundation for scalable, data-driven monitoring of cognitive demands in technology-mediated public health settings.

## Implications for Public Health

This work has implications for public health and clinical training environments, where cognitive overload, fatigue, and disengagement can compromise learning effectiveness and downstream performance. Traditional evaluations in these settings rely heavily on observable outcomes, such as test scores or task completion time, which may fail to capture underlying cognitive strain.

Our findings demonstrate that cognitive effort fluctuates systematically across task segments, even when performance remains relatively stable. This suggests that segment-level monitoring of cognitive effort can provide early indicators of overload or disengagement that are not visible through performance metrics alone. In public health training contexts such as workforce preparation, continuing education, or simulation-based assessment, this capability could support safer and more effective training design.

At the individual level, the ability to estimate cognitive effort using predicted performance further enhances scalability. In many real-world public health settings, immediate or reliable performance feedback may not be available. The proposed framework shows that cognitive effort can still be estimated objectively using brain signals and machine learning–derived proxies. This opens the possibility of adaptive digital training systems that respond to learners' cognitive states in real time, for example, by adjusting task difficulty, prompting rest, or reallocating instructional support.

Overall, this work provides a foundation for integrating objective cognitive effort monitoring into technology-mediated public health training systems, with potential benefits for reducing cognitive overload, preventing burnout, and supporting sustainable workforce performance.

### Limitations and Future Work

In our study, the sample size (N = 16) was sufficient for detecting group-level effects, but it limits the generalizability of machine learning models trained on high-dimensional fNIRS features. This constraint may explain the observed performance ceiling in score prediction. While participant-independent cross-validation was applied to reduce overfitting, future studies with larger and more diverse samples are needed to capture broader inter-individual variability.

Furthermore, the current analysis was conducted offline. Although this establishes the feasibility of the proposed framework, additional work is required to evaluate its computational efficiency and latency in real-time, closed-loop settings.

Future research will focus on scaling data collection to support more expressive modeling approaches and exploring multimodal sensor integration to improve robustness. From a system design perspective, personalization strategies such as individual baseline calibration may further enhance effort estimation.

A key distinction of this work is its emphasis on neural efficiency as a process rather than learning outcomes as an endpoint. While significant fluctuations in efficiency were observed across task segments, the relationship between real-time cognitive efficiency and long-term knowledge retention remains an open question. Addressing this gap may enable the development of adaptive training systems that not only assess performance but also monitor cognitive workload and fatigue—capabilities that are particularly relevant for public health and clinical workforce training environments.

## Conclusion

This paper proposed and validated a two-stage framework for estimating cognitive effort in digital learning tasks by combining fNIRS brain signals with machine learning–based performance prediction. In the first stage, we analyzed group-level cognitive effort using actual task performance and neural measures. The results showed statistically significant changes in relative neural efficiency and involvement across consecutive task segments, confirming that the task structure induced meaningful variations in cognitive effort beyond performance scores alone. In the second stage, we evaluated whether individual cognitive effort could be estimated using machine-learning–predicted performance instead of actual scores. Although performance prediction accuracy was moderate (67%), the resulting cognitive effort metrics closely preserved the distribution and segment-wise patterns observed with actual scores. This indicates that predicted performance is sufficient for estimating cognitive effort without substantially distorting efficiency and involvement measures. Together, these findings demonstrate the feasibility of estimating cognitive effort at both group and individual levels using brain signals and artificial intelligence. From a public health perspective, this framework provides a scalable foundation for objectively monitoring cognitive workload and mental fatigue in technology-mediated training and assessment environments, where direct performance feedback may be delayed, incomplete, or unavailable.

## Acknowledgment

We sincerely thank our lab members and all the study participants for their support and contribution. We also gratefully acknowledge the support of the National Science Foundation (NSF Award Nos. 2222661–2222663, 2321274, and 2426003). Any opinions, findings, and conclusions expressed in this material are those of the authors and do not necessarily reflect the views of the sponsors.

Ms. Sharmin may be contacted at shayla@udel.edu.